\documentstyle[12pt]{article}
\advance \topmargin by -\headheight
\advance \topmargin by -\headsep
     
\evensidemargin \oddsidemargin
\begin{document}
%               macros formatting and equations 
\topmargin -.6in
\def\br{\begin{eqnarray}}
\def\er{\end{eqnarray}}
\def\be{\begin{equation}}
\def\ee{\end{equation}}
\def\nn{\nonumber}
\def\({\left(}
\def\){\right)}
\def\a{\alpha}
\def\b{\beta}
\def\d{\delta}
\def\D{\Delta}
\def\eps{\epsilon}
\def\g{\gamma}
\def\G{\Gamma}
\def\h{ {1\over 2}  }
\def\hp{ {+{1\over 2}}  }
\def\hm{ {-{1\over 2}}  }
\def\k{\kappa}
\def\l{\lambda}
\def\L{\Lambda}
\def\m{\mu}
\def\n{\nu}
\def\o{\over}
\def\O{\Omega}
\def\p{\phi}
\def\rh{\rho}
\def\s{\sigma}
\def\t{\tau}
\def\th{\theta}
\def\ii {\'\i  }
\vskip .6in

\begin{center}
{\large {\bf Remark on Shape Invariant Potential }} 
\end{center}

\normalsize
\vskip .4in
\begin{center}
{Elso Drigo Filho\footnotemark\footnotetext{Work partially supported by CNPq
and FAPESP}} \\
\par \vskip .1in \noindent
Instituto de Bioci\^encias, Letras e Ci\^encias Exatas-UNESP\\
Departamento de F\ii sica \\
Rua Cristov\~ao Colombo, 2265\\
15055 S\~ao Jos\'e do Rio Preto, SP, Brazil\\
\par \vskip .3in

\end{center}

\begin{center}
{Regina Maria  Ricotta}
\par \vskip .1in \noindent
Faculdade de Tecnologia de S\~ao Paulo, CEETPS-UNESP\\
Pra\c ca Fernando Prestes, 30 \\
01121-060 S\~ao Paulo, SP, Brazil \\
\par \vskip .3in
 
\end{center}

\begin{center}
{\large {\bf ABSTRACT}}\\
The usual concept of shape invariance is discussed and one extension of this 
concept is suggested.
\end{center}
\par \vskip .3in
Gedenshtein \cite{Gedenshtein} defined the ``shape invariant" potentials by the
relationship
\be
\label{condition}
V_+(x; a_0) - V_-(x; a_1) = W^2(x,a_0) + W'(x, a_0) - W^2(x; a_1) + W'(x; a_1) =
R(a_1)               
\ee 
where $W(x; a)$ is the superpotential, $a_0$ and $a_1$ stand for parameters of 
the supersymmetric partner potentials $V_+$ and $V_-$, $R(a)$ is a constant. The 
supersymmetric partners are related with the supersymmetric Hamiltonian  in an
usual way \cite{Cooper}, $V_+ = W^2 - W'$ and $V_- = W^2 + W'$.
     
The relationship beetween shape invariance and solvable potentials  is
discussed by se\-ve\-ral authors (see, for instance, \cite{Levai} and
\cite{Cooper2}). Other  mathematical aspects of shape invariant potentials are
also present in the literature, for example in the supersymmetric WKB
appro\-xi\-ma\-tion, \cite{Dutt}, Berry  phase, \cite{Bhaumik}, and  in 
the path-integral  formulation, \cite{De}.
     
There is a general conclusion about these kind of potentials which is 
that the concept of shape invariance is a sufficient but not a necessary
condition for  the potential to become exactly solvable, \cite{Cooper2}. 
     
In a recent work, \cite{Drigo1},  the Hulth\'en potential was studied from the 
Supersymmetric Quantum Mechanics formalism. This potential has an interesting 
property, that is when the angular momentum is zero,  $l=0$,  it is not shape
invariant in the sense expressed in  ref.[1]. However, it is still possible to
construct a general form of the  potentials in the super-family of Hamiltonians:
\be
\label{potential}
V_n(r) - E_0^{(n)} = W_n^2(r) -{d\o dr}W_n(r) = 
{n(n-1){\d}^2 e^{-2\d r}\o 2(1-e^{-\d r})^2} -{[n(1-n)\d +2]\d e^{-\d r}\o
2(1-e^{-\d r})} + {1\o 2} (-{n\o 2}\d + {1\o n})^2.
\ee
where $n=1,2,3...$ labels the $n$-th member of the super-family whose
ground-state is $ E_0^{(n)}$, 
($n=1$ and $2$
correspond  to the two first members $V_+$ and $V_-$, respectively, except by
addictive constants in $V_-$)  
 and $\delta$  is a fixed parameter.  For $n=1$ the potential in (\ref{potential}) 
leads us to the usual Hulth\'en potential $V_H$
\be
\label{Potential}
V_+(r) = V_H(r) - E_0^{(1)} = - {\d e^{-\d r}\o 1-e^{-\d r}} + {1\o 2}({1-\d \o
2})^2, 
\ee
and from (\ref{potential}) it is easy to note that the condition (\ref{condition})
is not satisfied, i.e., $V_H$ is not shape invariant but the whole
super-family has the same functional form given by equation (\ref{potential}).
     
Taking the previous example, it is possible to suggest an extension
of the  concept of shape invariance.  This invariance would be associated
with the functional form of the whole super-family  potentials and not only with
the first two  members ($V_+$ and $V_-$), since all the members of super
family can be written in a general functional form in terms of one or more 
parameters (as the natural number  $n$ in Hulthen potential case) . In other
words, it is possible  to construct a general expression for all potentials of
the  super-family. 
     
The simple example of the free particle in a box can be used to make clear
the   above idea. The Hamiltonian $H$ in this case is
\be
\label{box}
H_+ = H - E_0^{(1)} = - {d^2\o d^2 x} - 1 \;;\;\;\;\; -{\pi \o 2} < x < +{\pi \o 2}
\ee
where the constant term $(-1)$ sets the eigenvalue of the ground state of $H_+$
to zero, \cite{Drigo2}. In this case the general form for the superpotential
is
\be        
W_n(x) = n\; tan(x) 
\ee
where $n$ is a natural number different from  zero, ($ n=1,2,3...$). The
super-family is such that $E_n^{(1)}=n^2$ and the $n$-th member  of the  
super-family potential is
\be                 
V_n(x) - E_0^{(n)} =  {n(n-1) \o cos^2(x)} - n^2.
\ee
    
Thus, it is not shape invariant in the Gedenshtein's sense, 
\cite{Gedenshtein}, since $V_+ = -1$ and
$V_- = {2 \o cos^2(x)}-1$, whereas it is shape invariant in the extended
sense. 
  
In our definition the potentials are shape invariant when it is possible to 
construct a  
super-family whose members have the same functional form. On the other hand, in 
the usual definition introduced by Gedenshtein, once relation (\ref{condition}) 
is
satisfied it is possible to find  all the members of the super-family. However, 
having built a super-family it does not necessarily mean that relation
(\ref{condition}) is satisfied, as shown in the two examples above of the
Hulth\'en potential and the particle in a box. In other 
words, Gedenshtein's condition of shape invariance is sufficient but not a 
necessary condition to obtain the super-family. 

The interesting  
question to be studied now is if the extended shape  invariance is a necessary
condition to the potential to be exactly solvable. Other questions concerning
shape invariance, \cite{Dutt},
\cite{Bhaumik}, \cite{De},  can also be analysed using this extended concept. 
\vskip 1cm
We acknowledge Dr. G. Levai for a critical reading of the manuscript.  
One of us (E.D.F.) would like to thank Prof. A. Inomata for calling his
attention to the non-shape invariance problem of the Hulth\'en potential.

\end{document}